\documentclass[twoside]{ae100prg}

\usepackage{graphicx}
\usepackage[breaklinks]{hyperref}
\usepackage{booktabs}

\usepackage{amsbsy}
\usepackage{amsmath}
\usepackage{amsthm}

\newcommand{\dm}{n}
\newcommand{\hook}{\raisebox{-0.35ex}{\makebox[0.6em][r]
{\scriptsize $-$}}\hspace{-0.15em}\raisebox{0.25ex}{\makebox[0.4em][l]{\tiny
 $|$}}}

\newcommand{\be}{\begin{equation}}
\newcommand{\ee}{\end{equation}}
\newcommand{\ba}{\begin{eqnarray}}
\newcommand{\ea}{\end{eqnarray}}
\newcommand{\cwedge}[1]{\mathop{\wedge}_{{}^{#1}} }
\newcommand{\nn}{\nonumber}

\newcommand{\even}{{\mathrm{e}}}
\newcommand{\odd}{{\mathrm{o}}}
\newcommand{\eps}{\varepsilon}
\newcommand{\A}[1]{A^{\!(#1)}}
\newcommand{\cv}[1]{{\partial}_{#1}}
\newcommand{\KV}[1]{\xi^{(#1)}}
\newcommand{\psc}[1]{\Psi_{#1}}
\newcommand{\lst}[1]{\langle{#1}\rangle}
\newcommand{\psf}[1]{\tilde\Psi_{#1}}

\begin{document}
\title{Hidden Symmetries of the Dirac equation in curved space-time}


\author{Marco Cariglia$^1$} 

\address{$^1$ Universidade Federal de Ouro Preto, ICEB, Departamento de F\'isica.
  Campus Morro do Cruzeiro, 35400-000 - Ouro Preto, MG - Brasil}

\email{marco@iceb.ufop.br}

\begin{abstract}
These are introductory notes on the study of the Dirac equation in curved spacetime and its relation to hidden symmetries of the dynamics. We present general results on the relation between special spacetime tensors and hidden symmetries, both for the full Dirac equation and for its semi-classical limit, the spinning particle. A concrete application of the general results is provided by the case of rotating higher dimensional black holes with cosmological constant, which we discuss. For these metrics the Dirac equation is separable and the relation between this and hidden symmetries is explained. 
\end{abstract}

\section{Introduction}
The Dirac equation, since its derivation in 1928, has successfully described the relativistic hydrogen atom and phenomena such as the existence of anti-particles. A natural evolution of the theory, stimulated by the progress in General Relativity, is that of studying relativistic spin $\frac{1}{2}$ particles on a curved background, such as for example the Schwarzschild and Kerr spacetimes. Further development in unification theories such as String/M--Theory and in cosmological models lead to the additional ingredient of considering extensions of General Relativity and quantum field theories to higher dimensions than four. 
 
Parallel to this, the natural interest of physicists in solutions of the Dirac equation that can be obtained by separation of variables lead to the mathematical problem of finding a theory of separation of variables for this system of first order equations. In the case of the classical Hamilton-Jacobi equations, the Schr\"{o}dinger equation and the Klein Gordon equation there exists a well understood theory of separation of variables \cite{Benenti1997,BenentiFrancaviglia1979,FrancavigliaDemianski1980,KalninsMiller1981,FrancavigliaMcLenaghan2002}. In this theory the main objects playing a role are special tensors, namely Killing vectors and rank 2 Killing-St\"{a}ckel tensors. With these one can build either conserved quantities in the classical theory, or symmetry operators in the quantum mechanical one. In the case of the Dirac equation instead a complete theory of separation of variables is still lacking. Several known cases of separation of variables for the Dirac equation involve only symmetry operators of first order in the derivatives, such as the  Dirac equation in the Kerr metric or the higher dimensional Kerr-NUT-(A)dS metrics that we will discuss in this review. First order symmetry operators have been built in 4 dimensions first and successively in arbitrary dimensions and signature \cite{MclenaghanSpindel1979,CarterMcLenaghan1979,KamranMcLenaghan1984,BennCharlton1997,BennKress2004}. 
However, Fels and Kamran have shown  that there exist cases where the Dirac equation is separable but the separability is underlain by the existence of symmetry operators of order higher than one. In some cases second order symmetry operators have been built, see \cite{McLenaghanRastelli2011} and references therein, but neither the general construction is  known for arbitrary dimension, nor there is control on necessary and sufficient conditions for separability. 
 
In this review we present the current knowledge on first order symmetry operators of the Dirac equation. We show how these are in exact correspondence with special Conformal Killing-Yano tensors, and the relation between spacetime differential forms and Clifford algebra valued operators. We discuss one important application, the separability of the Dirac equation in Kerr-NUT-(A)dS metrics, which is fully accounted for by a complete set of linear symmetry operators that are mutually commuting and admit common separable spinorial eigenfunctions. We also briefly discuss the semi-classical limit of the Dirac equation, the theory of the supersymmetric spinning particle. In this theory the analogue of linear symmetry operators is given by phase space functions that are linear in the momenta and correspond to generators of extra supersymmetries. For Kerr-NUT-(A)dS metrics it is possible to show that the bosonic sector of the theory is integrable, and its integrability is related to the presence of Killing vectors and of a set of new conserved quantities that are quadratic in the momenta and generalise the quadratic conserved quantities of a scalar particle to the case with spin.

\section{Gamma matrices and differential forms}
In this section we discuss the one-to-one map between differential forms on a spin manifold and sections of its Clifford bundle. The application of interest for this review is the fact that the properties of Conformal Killing-Yano tensors, which are differential forms, automatically lift to those of appropriate differential operators defined on the Clifford bundle. 

We model spacetime as a (pseudo-)Riemannian spin manifold $M$ of dimension ${\dm}$ with metric ${g_{\mu\nu}}$ and local coordinates $\{ x^\mu \}$. We use lowercase Greek indices to denote 'curvy' components of spacetime tensors, associated to general diffeomorphism transformations, and lowercase latin indices to denote 'flat' components, associated to $SO(\dm)$ or $SO(1, \dm -1)$ transformations. Each fiber of the Clifford bundle has structure of the Clifford algebra generated by the gamma matrices ${\gamma^\mu}$, which connect the Clifford bundle with the tangent space. The gamma matrices satisfy 
\begin{equation}\label{ggmetric}
    \gamma^\mu\,\gamma^\nu + \gamma^\nu\, \gamma^\mu = 2g^{\mu\nu}\;,
\end{equation}
which allows to reduce any element $\slash\mspace{-10mu}\alpha$ of the Clifford algebra to a sum of antisymmetric products ${\gamma^{\mu_1\dots \mu_p} :=\gamma^{[\mu_1}}\dots\gamma^{\mu_p]}$ with appropriate coefficients: 
\begin{equation}\label{clobrepr}
    \slash\mspace{-10mu}\alpha
     = \sum_p \frac1{p!}\, \alpha^{(p)}_{\mu_1\dots \mu_p} \gamma^{\mu_1\dots \mu_p}\;.
\end{equation}
This representation is unique and the coefficients are given by anti-symmetric forms $\alpha^{(p)}_{\mu_1\dots \mu_p} \in \Omega^{(p)}(M)$, thus providing an isomorphism ${\gamma_*}$ of the Clifford bundle with the exterior algebra ${\Omega(M)=\bigoplus_{p=0}^{\dm} \Omega^p(M)}$ of inhomogeneous antisymmetric forms: $\slash\mspace{-10mu}\alpha = \gamma_* \alpha$, where $\alpha = \sum_p  \alpha^{(p)}$ is an inhomogeneous form. In the rest of the review, whenever the context makes it clear we will write $\alpha$ instead of $\slash\mspace{-10mu}\alpha$ to describe an element of the Clifford algebra, with the action of the isomorphism implied.

The metric allows to raise and lower indices (musical isomorphism): if $\alpha$ is a 1--form and $v$ a vector, we denote the corresponding vector and 1--form as ${\alpha^\sharp}$ and ${v^\flat}$, respectively. This generalises to higher rank tensors.
We define two operations on $\Omega(M)$. The `hook' operation (\emph{inner derivative}) is an action of a vector ${v}$ on any antisymmetric form ${\alpha}$. In components: 
\begin{equation}\label{hook}
    (v\hook\alpha)_{a_1\dots a_{p{-}1}}=v^b\alpha_{b a_1\dots a_{p{-}1}}\;.
\end{equation}
For a scalar ${\varphi}$, we set ${v\hook\varphi=0}$. The second operation is the wedge product. When it acts on a $p$--form $\alpha$ and a $q$--form $\beta$ it is defined so that in components ${(\alpha\wedge\beta)_{a\dots b\dots} = \frac{(p+q)!}{p!\,q!}\,\alpha_{[a\dots}\,\beta_{b\dots]}}$. 
 
The Clifford algebra relation \eqref{ggmetric} means that a product of any two rank $p$ and $q$ gamma matrices $\gamma^{\mu_1\dots \mu_p}$ and $\gamma^{\nu_1\dots \nu_q}$ matrices can be decomposed in terms of other gamma matrices. In particular it can be shown that for $\alpha\in\Omega^p(M)$,  $\beta\in\Omega^q(M)$ Clifford bundle forms, with $p\leq q$, the Clifford product expands as
\be
\alpha \beta = \sum_{m=0}^p \frac{ (-1)^{m(p-m) + [m/2]}}{m!} \alpha \cwedge{m}  \beta  \, , \label{usefulProduct1}
\ee
where the $\cwedge{m}$ contraction operator is defined recursively as 
\ba
\alpha\cwedge{0}\beta &=& \alpha \wedge \beta  \, , \nn \\
\alpha\cwedge{k}\beta &=& (X_a \hook \alpha ) \cwedge{k-1} (X^a \hook \beta) \qquad (k\ge 1) \, , \nn \\
\alpha\cwedge{k}\beta &=& 0 \qquad\qquad\qquad\qquad\qquad\, (k<0) \, .\label{cwedgedef}
\ea

Given a set of $\dm$--beins $\left\{ e^a_\mu \right\}$, we can build $\dm$ 1--forms $e^a = e^a_\mu dx^\mu$, with $X^a = (e^a)^\sharp$ a dual vector basis. The $e^a$ are mapped under $\gamma_*$ to 'flat' gammas matrices that satisfy \eqref{ggmetric} with the flat metric $\eta^{ab}$ instead of $g$. Flat and curvy indices can be transformed one into the other using either $e^a_\mu$ or its inverse $E_a^\mu$, such that $e^a_\mu E^\mu_b = \delta^a_b$ and $e^a_\mu E_a^\nu = \delta^\nu_\mu$.  We can also group the $e^a$ together and consider a single object $\gamma_* (e^a)$ which transforms as an $SO(\dm)$ tensor. We lift the covariant derivative given on $\Omega(M)$ to one on $\gamma_* \left( \Omega(M) \right)$ by asking that for any $\alpha$ in the Clifford bundle 
\begin{equation}\label{covdervielbein}
    \nabla_{\!a}\alpha = \partial_a \alpha - \omega{}_a \cwedge{1} \alpha \;, 
\end{equation}
where $\partial_a \alpha = X_a  [\alpha] = E_a^\mu \partial_\mu \alpha$, and  $\omega_a$ is the connection 2-form ${\omega_a= \frac12\omega_{abc} e^b\wedge e^c}$ and $\omega_{abc}$ are the components of the spin connection. For a form which is also an $SO(\dm)$ tensor such as $e^a$ the covariant derivative becomes 
\begin{equation}\label{nablae=0invielbein}
    \nabla_{\!a} e^b = \partial_a e^b + \omega{}_a{}^b{}_c\, e^c - \omega_a\cwedge{1} e^b \;. 
\end{equation}
In particular for the $\dm$--bein tensor itself $\nabla_a e^b = 0$. 

Lastly we introduce the degree operator ${\pi}$ that acts on an inhomogeneous form $\alpha = \sum_{p} \alpha^{(p)}$ as $
\pi \alpha = \sum_{p=0} p \, \alpha^{(p)}$. 

In this formalism the Dirac operator is written as ${D\equiv e^a\nabla_{\!a}=\nabla_{\!a} e^a}$, the exterior derivative acting on forms as ${d = e^a \wedge \nabla_{\!a} = \nabla_{\!a}\, e^a \wedge}$, and the co-differential  ${\delta = - X^a \hook \nabla_{\!a} = -\nabla_{\!a}\, X^a \hook}$. All these expressions are to be understood as operators acting on the right.

\section{Special Killing-Yano tensors\label{eq:KY_tensors}} 
{\it Conformal Killing-Yano} tensors (CKY) are forms $f_{\mu_1 \dots \mu_p} = f_{[\mu_1 \dots \mu_p]}$ such that 
\be 
\nabla_\lambda \, f_{\mu_1 \dots \mu_p} = \nabla_{[\lambda} \,  f_{\mu_1 \dots \mu_p]} + \frac{p}{D-p+1} g_{\lambda [\mu_1} \nabla^\rho f_{|\rho| \mu_2 \dots \mu_p]} \, , 
\ee
or equivalently without using components 
\be \label{eq:CKY_definition}
\nabla_X f = \frac{1}{\pi +1} X \hook d f - \frac{1}{\dm - \pi +1} X^\flat \wedge \delta f \, , 
\ee
for any vector $X$. When $p=1$ this reduces to the Killing equation. 

The formula above generalises automatically to the case of inhomogeneous forms. When $f$ is co-closed, $\delta f = 0$, $f$ is called a Killing-Yano form (KY), and when it is closed, $df =0$, it is called a closed conformal Killing-Yano form (CCKY). Equation \eqref{eq:CKY_definition} is invariant under Hodge duality, interchanging KY and CCKY tensors. 
A symmetry operator for the Dirac equation is an operator $S$ that R--commutes with $D$, or in other words such that $DS=RD$ for some operator ${R}$. It maps solutions of the massless Dirac equation into solutions. Benn, Charlton, and Kress \cite{BennCharlton1997, BennKress2004} have shown the important result that, in all dimensions  $\dm$ and arbitrary signature, R--symmetry operators of the massless Dirac operator that are first-order in derivatives are in one to one correspondence with CKY forms. Any such operator $S$ can be written as 
\be
S=S_f+\alpha D\,,
\ee
where $\alpha$ is an arbitrary inhomogeneous form, and $S_f$ is given in terms of an inhomogeneous CKY form $f$ obeying \eqref{eq:CKY_definition} according to 
\be\label{SOProp1}
S_f=X^a\hook f\, \nabla_a+\frac{\pi-1}{2\pi}d f-\frac{n-\pi-1}{2(n-\pi)}\delta f\;.
\ee
The freedom of adding an arbitrary form $\alpha$ is unavoidable. In \cite{MarcoDavidPavel2011} it was shown that, as a special case, the most general first-order operator $S$ that strictly commutes with $D$ splits into the Clifford even and Clifford odd parts
\be\label{Scom}
  S=S_\mathrm{e}+S_\mathrm{o}\,,
\ee
where
\begin{align}
    S_\mathrm{e}&= K_{\omega_\odd} \equiv X^a\hook\omega_\odd\nabla_{\!a}
       + \frac{\pi-1}{2\pi}d\omega_\odd\;,
       \label{Kdef}\\
    S_\mathrm{o}&= M_{\omega_\even} \equiv e^a\wedge\omega_\even\nabla_{\!a}
       - \frac{\dm-\pi-1}{2(\dm-\pi)}\delta\omega_\even\, , \label{Mdef}
\end{align}
where ${\omega_\odd}$ is an odd KY form and $\omega_\even$ is an even CCKY form. 

Given a KY $p$--form $f$ it is possible to see that the tensor 
\be 
K^{\mu\nu} = f^\mu {}_{\lambda_1 \dots \lambda_{p-1}} f^{\nu \lambda_1 \dots \lambda_{p-1}} \, ,
\ee
is Killing--St\"{a}ckel. Such a tensor  satisfies $\nabla^{( \lambda} K^{\mu_1 \dots \mu_p)} = 0$ and is associated to conserved quantities of higher order in the momenta for the theory of the classical particle. 

In \cite{DavidPavelValeriDonPage2007} it was shown that CCKY tensors form an algebra under the wedge product. In particular, closed conformal Killing--Yano tensors of rank 2 that are non-degenerate are called \textit{Principal conformal Killing--Yano (PCKY) tensors}. They are crucial for the integrability of various systems in four and higher dimensional black hole spacetimes. 

Generalisations of these concepts to the case of metrics with torsion and fluxes have been treated in \cite{DavidClaudeHouriYasui2010,DavidPavelClaude2010}.

\section{Kerr-NUT-(A)dS black holes} 
While a classification of Lorentzian metrics with a PCKY tensor is not available, the analogue problem in Riemannian signature has been solved \cite{HouriOotaYasui2007,KrtousFrolovKubiznak2008}. The most general {\em canonical metric} admitting a PCKY tensor in $\dm = 2N + \eps$ dimensions, $\eps=0,1$, is given by 
\be  \label{metric}
ds^2
  = \sum_{\mu=1}^N\biggl[ \frac{d x_{\mu}^{\;\,2}}{Q_\mu}
  +Q_\mu\Bigl(\,\sum_{j=0}^{N-1} \A{j}_{\mu}d\psi_j \Bigr)^{\!2}  \biggr]  + \eps S \Bigl(\,\sum_{j=0}^N \A{j}d\psi_j \Bigr)^{\!2}.
\ee
Here, coordinates $x_\mu\, (\mu=1,\dots,N)$ stand for the (Wick rotated) radial coordinate and longitudinal angles, and Killing
coordinates $\psi_k\; (k=0,\dots,N-1 +\eps)$ denote time and azimuthal angles associated with Killing vectors
${\KV{k}} =\cv{\psi_k}$. We have further defined the functions 
\ba
Q_\mu&=&\frac{X_\mu}{U_\mu}\,,\quad U_{\mu}=\prod\limits_{\nu\ne\mu} (x_{\nu}^2-x_{\mu}^2)  \;,\quad S = \frac{-c}{\A{N}} \, ,\label{eq:UandS_def}\\
\A{k}_{\mu}&=&\hspace{-5mm}\!\!
    \sum\limits_{\substack{\nu_1,\dots,\nu_k\\\nu_1<\dots<\nu_k,\;\nu_i\ne\mu}}\!\!\!\!\!\!\!\!\!\!
    x^2_{\nu_1}\cdots\, x^2_{\nu_k}\;,\ \
\A{k} = \hspace{-5mm} \sum\limits_{\substack{\nu_1,\dots,\nu_k\\\nu_1<\dots<\nu_k}}\!\!\!\!\!\!
    x^2_{\nu_1}\cdots\, x^2_{\nu_k}\; .\label{eq:A_def}\quad
\ea
The quantities ${X_\mu}$ are functions of a single variable ${x_\mu}$, and $c$ is an arbitrary constant.
The vacuum (with a cosmological constant) black hole geometry is recovered by setting
\begin{equation}\label{BHXs}
  X_\mu = \sum_{k=\eps}^{N}\, c_{k}\, x_\mu^{2k} - 2 b_\mu\, x_\mu^{1-\eps} + \frac{\eps c}{x_\mu^2} \; .
\end{equation}
This choice of $X_\mu$ describes the most general known Kerr-NUT-(A)dS spacetimes in all dimensions \cite{ChenLiuPope2006}. The constant $c_N$ is proportional to the cosmological constant and the remaining constants are related to angular momenta,
mass and NUT parameters. 

The PCKY tensor reads \cite{DavidFrolov2006}
\be
h=db\,,\quad b=\frac{1}{2}\sum_{j=0}^{N-1}A^{(j+1)}d\psi_j\,.
\ee 
The $2j$-forms $h^{(j)}$, which are the $j$-th wedge power of the PCKY tensor $h$, $h^{(j)} = h\wedge \dots \wedge h$, form the tower of associated closed conformal Killing--Yano tensors. Their Hodge duals are KY forms and can be `squared' to rank 2 Killing--St\"{a}ckel tensors.

\section{Separability of the Dirac equation in the Kerr-NUT-(A)dS metric} 
According to the results of section \ref{eq:KY_tensors}, to the $N+\eps$ Killing vectors $\xi^{(0)}, \dots, \xi^{(N-1+\eps)}$ are associated the operators $K_{\xi^{(0)}}, \dots, K_{\xi^{(N-1+\eps)}}$, which commute with the Dirac operator $D$. And to the $N$ CCKY forms $h^{(j)}$ are associated operators $M_{h^{(1)}}, \dots, M_{h^{(N-1)}}$, which also commute with $D$. In \cite{MarcoDavidPavel2011} it has been shown that all these operators are in fact mutually commuting. Thus it is possible to diagonalise them simultaneously and to look for common spinorial eigenfunctions. 

The Dirac equation in this metric had been shown to be separable by a direct calculation in \cite{OotaYasui2008}. A geometrical understanding of the result has been given in \cite{MarcoDavidPavel2011_2}, where it has been shown that the solution to the eigenvalue problem 
\be\label{eigenvaluetilde}
K_{\xi^{(k)}} \chi = i\,\psc{k}\chi\;,\quad M_j \chi = m_j \chi\;,
\ee
can be found in the tensorial R--separated form
\begin{equation}\label{tens_sep}
    \chi = R \exp\bigl({\textstyle i\sum_k\psc{k}\psi_{k}}\bigr)\,
           \bigotimes_\nu \chi_\nu\;,
\end{equation} 
where $\left\{\chi_\nu \right\}$ is an $N$-tuple of 2-dimensional spinors and ${R}$ is an appropriate Clifford bundle-valued prefactor. $\chi_\nu$ depends only on the variable ${x_\nu}$, $\chi_\nu=\chi_\nu(x_\nu)$, and satisfies the equation 
\begin{equation}\label{chieq}
\begin{split}
&\Biggl[\Bigl(
    \frac{d}{dx_\nu}+\frac{X_\nu'}{4X_\nu}
    +\frac{\tilde \Psi_\nu}{X_\nu}\iota_{\lst{\nu}}
    +\frac{\eps}{2x_\nu}\Bigr)\,\sigma_{\lst{\nu}} \\
&\mspace{70mu}- \,\frac{\bigl(- \iota_{\lst{\nu}} \bigr)^{\!N\!-\!\nu}}{\sqrt{|X_\nu|}}
   \Bigl(\eps \frac{i\sqrt{-c}}{2x_\nu^2} +m_\nu\Bigr)\Biggr]\,\chi_\nu=0\, , 
\end{split}
\end{equation}
where 
\begin{equation}\label{psfdef}
    \psf{\mu} = \sum_k \psc{k} (-x_\mu^2)^{N{-}1{-}k}\; , 
\end{equation}
and 
\be  \label{eq:chf_mu_definition}
m_\nu =  \sum_{j} (-i)^j m_j \left( -\iota_{\lst{\nu}} x_\nu \right)^{N-1-j} \, .
\ee
 The operator $\iota_{\lst{\nu}}$ acts as a $\sigma_3$ Pauli matrix on the 2-spinor $\chi_\nu$ while leaving the other spinors $\chi_\mu$, $\mu \neq \nu$, invariant, and similarly $\sigma_{\lst{\nu}}$ acts as as $\sigma_1$. 
 
The solution \eqref{tens_sep} is the same as that given in \cite{OotaYasui2008}. The arbitrary integration constants found there are related to the eigenvalues $\psc{k}$ and $m_j$.

\section{The spinning particle} 
The spinning particle theory can be thought of as a semi-classical description of a Dirac fermion. The degrees of freedom are coordinates $x^\mu$ and Grassmannian variables $\theta^a$ related to the spin. The Hamiltonian is given by 
\ba
H=\frac{1}{2}\Pi_\mu\Pi_\nu g^{\mu\nu}\,,\ \ 
\Pi_\mu=p_\mu - \frac{i}{2}\theta^a\theta^b\omega_{\mu ab}=g_{\mu\nu}\dot x^\nu\,,\qquad
\ea
where $p_\mu$ is the momentum canonically conjugate to $x^\mu$ and $\Pi_\mu$ is the covariant momentum. Poisson brackets are defined as 
\be\label{brackets}
\{F,G\}=\frac{\partial F}{\partial x^\mu}\frac{\partial G}{\partial p_\mu}-
\frac{\partial F}{\partial p_\mu}\frac{\partial G}{\partial x^\mu}+
i(-1)^{a_F}\frac{\partial F}{\partial \theta^a}\frac{\partial G}{\partial \theta_a}\,,
\ee
where $a_F$ is the Grassmann parity of $F$.
 
The theory is worldsheet supersymmetric and the generator of supersymmetry is given by 
\be\label{Qdef}
Q=\theta^a e_a{}^\alpha \Pi_\alpha\,,
\ee
which obeys
\be
\{H,Q\}=0\,, \quad \{Q,Q\}=-2iH\,.
\ee

Equations of motion are accompanied by two physical (gauge) conditions
\be\label{gaugecond}
2H=-1\,,\quad Q=0\,,
\ee

In the Kerr-NUT-(A)dS spacetimes there are $(N+\varepsilon)$ Killing vectors $\xi_{(k)}$. It is possible to show that with these one can construct  bosonic superinvariants linear in velocities, given by 
\be\label{Qk}
Q_{\xi_{(k)}}=\xi_{(k)}^\alpha\Pi_\alpha-\frac{i}{4}\theta^a\theta^b (d\xi_{(k)})_{ab}\,.
\ee 
These can be used to express some components of the velocities $\Pi$ in terms of the conserved quantities and of the $\theta$ variables. In \cite{MarcoDavid2012} it was shown that it is possible to find $N$ further bosonic supersymmetric conserved quantities ${\cal K}_{(j)}$, this time quadratic in the velocities. These new quantities will not be conserved nor supersymmetric in a general metric, but they are for Kerr-NUT-(A)dS. The $(N+\eps) + N = \dm$ quantities are all independent and using them it is possible to express all the components of $\Pi$, thus showing that the bosonic sector of the theory is integrable. 
 
The quantities ${\cal K}_{(j)}$ are written as 
\ba\label{quadr}
{\cal K}_{(j)}&=&K_{(j)}^{\mu\nu} \Pi_\mu\Pi_\nu+{\cal L}_{(j)}^\mu\Pi_\mu+{\cal M}_{(j)}\,,\nonumber\\
{\cal L}^\mu_{(j)}&=&\theta^{a}\theta^b L_{(j) ab}{}^\mu\,,\quad  
{\cal M}_{(j)}=\theta^{a}\theta^b\theta^c\theta^d M_{(j)abcd}\,.\quad
\ea
The tensors $K$, $L$ and $M$ are given by 
\ba\label{solution}
K^{\mu\nu}&=&f^{\mu \kappa_1\dots \kappa_{p-1}}f^\nu{}_{\kappa_1\dots \kappa_{p-1}}\,,\label{solK}\nonumber\\
L_{\mu\nu}{}^\rho&=& -\frac{2i}{p+1} f_{[\mu|\kappa_1\dots \kappa_{p-1}|}(df)_{\nu]}{}^{\rho\kappa_1\dots \kappa_{p-1}}\nonumber\\
&-&\vspace{0.5cm} \frac{2i}{p+1} (df)_{\mu\nu \kappa_1\dots \kappa_{p-1}}f^{\rho\kappa_1\dots \kappa_{p-1}}\,,\qquad \label{solL}\\
M_{\mu\nu\rho\sigma}&=&-\frac{i}{4} \nabla_{[\mu} L_{\nu\rho\sigma]}\,.\nonumber\label{solM}
\ea
Here $f_{\mu_1\dots \mu_p}$ is the appropriate rank-$p$ Killing--Yano tensor present in the spacetime.

\vspace*{1ex}
\section*{Acknowledgments}
The author is partially funded by Fapemig under the project CEX APQ 2324-11.

\section*{References}
\bibliography{ae100prg_Contribution_Cariglia}

\begin{thebibliography}{10}

\bibitem{Benenti1997}
Benenti, S., ``Intrinsic characterization of the variable separation in the
  Hamilton Jacobi equation'', {\em J. Math. Phys.}, {\bf 38}, 6578--6602,
  (1997).

\bibitem{BenentiFrancaviglia1979}
Benenti, S.  and Francaviglia, M., ``Remarks on certain separability structures
  and their applications to general relativity'', {\em Gen. Rel. and Grav.},
  {\bf 10}, 79--92, (1979).

\bibitem{BennCharlton1997}
Benn, I.~M.  and Charlton, P., ``{Dirac symmetry operators from conformal
  Killing-Yano tensors}'', {\em Class.Quant.Grav.}, {\bf 14}, 1037--1042,
  (1997).
  {\small[\href{http://arxiv.org/abs/gr-qc/9612011}{{arXiv:gr-qc/9612011}}]}.

\bibitem{BennKress2004}
Benn, I.~M.  and Kress, J.~M., ``First-order Dirac symmetry operators'', {\em
  Class. Quant. Grav.}, {\bf 21}(2), 427, (2004).

\bibitem{MarcoDavidPavel2011}
Cariglia, M., Krtous, P.  and Kubiznak, D., ``{Commuting symmetry operators of
  the Dirac equation, Killing-Yano and Schouten-Nijenhuis brackets}'', {\em
  Phys.Rev.}, {\bf D84}, 024004, (2011).
  {\small[\href{http://arxiv.org/abs/1102.4501}{{arXiv:1102.4501}}]}.

\bibitem{MarcoDavidPavel2011_2}
Cariglia, M., Krtou\v{s}, P.  and Kubiz\v{n}\'ak, D., ``{Dirac Equation in
  Kerr-NUT-(A)dS Spacetimes: Intrinsic Characterization of Separability in All
  Dimensions}'', {\em Phys.Rev.}, {\bf D84}, 024008, (2011).
  {\small[\href{http://arxiv.org/abs/1104.4123}{{arXiv:1104.4123}}]}.

\bibitem{McLenaghanRastelli2011}
Carignano, A., Fatibene, L., McLenaghan, R.~G.  and Rastelli, G., ``{Symmetry
  Operators and Separation of Variables for Dirac's Equation on Two-Dimensional
  Spin Manifolds}'', {\em SIGMA}, {\bf 7}, 57--69, (2011).
  {\small[\href{http://arxiv.org/abs/1102.0065}{{arXiv:1102.0065}}]}.

\bibitem{CarterMcLenaghan1979}
Carter, B.  and Mclenaghan, R.~G., ``Generalized Total Angular Momentum
  Operator For The Dirac Equation In Curved Space-time'', {\em Phys.Rev.}, {\bf
  D19}, 1093--1097, (1979).

\bibitem{ChenLiuPope2006}
Chen, W., L{\"u}, H.  and Pope, C.~N., ``{General Kerr NUT AdS metrics in all
  dimensions}'', {\em Class.Quant.Grav.}, {\bf 23}, 5323--5340, (2006).
  {\small[\href{http://arxiv.org/abs/arXiv:hep-th/0604125}{{arXiv:hep-th/06041%
25}}]}.

\bibitem{FrancavigliaDemianski1980}
Demianski, M.  and Francaviglia, M., ``Separability structures and Killing-Yano
  tensors in vacuum type-D space-times without acceleration'', {\em Int. J.
  Theor. Phys.}, {\bf 19}, 675--680, (1980).

\bibitem{FrancavigliaMcLenaghan2002}
Fatibene, L., Ferraris, M., Francaviglia, M.  and McLenaghan, R.G.,
  ``Generalized symmetries in mechanics and field theories'', {\em J. Math.
  Phys.}, {\bf 43}, 3147--3161, (2002).

\bibitem{DavidClaudeHouriYasui2010}
Houri, T., Kubiz\v{n}\'ak, D., Warnick, C.  and Yasui, Y., ``{Symmetries of the
  Dirac Operator with Skew-Symmetric Torsion}'', {\em Class.Quant.Grav.}, {\bf
  27}, 185019, (2010).
  {\small[\href{http://arxiv.org/abs/1002.3616}{{arXiv:1002.3616}}]}.

\bibitem{HouriOotaYasui2007}
Houri, T., Oota, T.  and Yasui, Y., ``Closed conformal Killing Yano tensor and
  Kerr NUT de Sitter space time uniqueness'', {\em Phys. Lett. B}, {\bf 656},
  214--216, (2007).
  {\small[\href{http://arxiv.org/abs/0708.1368}{{arXiv:0708.1368}}]}.

\bibitem{KalninsMiller1981}
Kalnins, E.~G.  and Miller, W., ``Killing Tensors and Nonorthogonal Variable
  Separation for Hamilton Jacobi Equations'', {\em SIAM J. Math. Anal.}, {\bf
  12}, 617--629, (1981).

\bibitem{KamranMcLenaghan1984}
Kamran, N.  and Mclenaghan, R.~G., ``Symmetry Operators For Neutrino And Dirac
  Fields On Curved Space-time'', {\em Phys.Rev.}, {\bf D30}, 357--362, (1984).

\bibitem{KrtousFrolovKubiznak2008}
Krtou\v{s}, P., Frolov, V.~P.  and Kubiz\v{n}\'ak, D., ``Hidden symmetries of
  higher-dimensional black holes and uniqueness of the Kerr-NUT-(A)dS
  spacetime'', {\em Phys. Rev. D.}, {\bf 78}(6), 064022, (2008).
  {\small[\href{http://arxiv.org/abs/0804.4705}{{arXiv:0804.4705}}]}.

\bibitem{DavidPavelValeriDonPage2007}
Krtou\v{s}, P., Kubiz\v{n}\'ak, D., Page, D.~N.  and Frolov, V.~P.,
  ``Killing-Yano tensors, rank-2 Killing tensors, and conserved quantities in
  higher dimensions'', {\em JHEP}, {\bf 2007}(02), 004, (2007).
  {\small[\href{http://arxiv.org/abs/hep-th/0612029}{{arXiv:hep-th/0612029}}]}.

\bibitem{MarcoDavid2012}
Kubiz\v{n}\'ak, D.  and Cariglia, M., ``{On Integrability of spinning particle
  motion in higher-dimensional black hole spacetimes}'', {\em Phys.Rev.Lett.},
  {\bf 108}, 051104, (2012).
  {\small[\href{http://arxiv.org/abs/1110.0495}{{arXiv:1110.0495}}]}.

\bibitem{DavidFrolov2006}
Kubiz\v{n}\'ak, D.  and Frolov, V.~P., ``{The hidden symmetry of higher
  dimensional Kerr NUT AdS spacetimes}'', {\em Class.Quant.Grav.}, {\bf 24},
  F1--F6, (2007).
  {\small[\href{http://arxiv.org/abs/arXiv:gr-qc/0610144}{{arXiv:gr-qc/0610144%
}}]}.

\bibitem{DavidPavelClaude2010}
Kubiz\v{n}\'ak, D., Warnick, C.  and Krtou\v{s}, P., ``{Hidden symmetry in the
  presence of fluxes}'', {\em Nucl.Phys.}, {\bf B844}, 185--198, (2011).
  {\small[\href{http://arxiv.org/abs/1009.2767}{{arXiv:1009.2767}}]}.

\bibitem{MclenaghanSpindel1979}
Mclenaghan, R.~G.  and Spindel, P., ``{Quantum Numbers For Dirac Spinor Fields
  On A Curved Space-time}'', {\em Phys.Rev.}, {\bf D20}, 409--413, (1979).

\bibitem{OotaYasui2008}
Oota, T.  and Yasui, Y., ``{Separability of Dirac equation in higher
  dimensional Kerr NUT de Sitter spacetime}'', {\em Phys. Lett. B}, {\bf 659},
  688--693, (2008).
  {\small[\href{http://arxiv.org/abs/0711.0078}{{arXiv:0711.0078}}]}.

\end{thebibliography}

\end{document}